\documentclass[aps,prl,twocolumn,superscriptaddress]{revtex4-1}
\usepackage{amsmath}
\usepackage{graphicx}
\usepackage{mathtools}
\usepackage{mathtools}
\usepackage{float}
\usepackage{empheq}
\usepackage{soul}
\usepackage{epstopdf}
\usepackage{bigints}
\usepackage{epstopdf}
\usepackage{amssymb}
\usepackage{wrapfig}
\usepackage{hyperref}
\usepackage{soul}
\usepackage{textcomp}
\usepackage[makeroom]{cancel}

\usepackage{float,graphicx}
\usepackage[usenames, dvipsnames]{color}

\hypersetup{
	colorlinks,
	citecolor=blue,
	filecolor=red,
	linkcolor=blue,
	urlcolor=blue,
	hyperfigures
}

\def\XXint#1#2#3{{\setbox0=\hbox{$#1{#2#3}{\int}$}
		\vcenter{\hbox{$#2#3$}}\kern-.5\wd0}}

\usepackage[page]{appendix}

\graphicspath{{./Figures/}}

\newcommand{\bv}{{\bf v}}

\newcommand{\bx}{{\bf x}}

\newcommand{\bn}{\mathbf n}

\newcommand{\siga}{\boldsymbol{\sigma}^{\text{act}}}
\newcommand{\sige}{\boldsymbol{\sigma}^{\text{el}}}
\newcommand{\pll}{\parallel}
\newcommand{\md}{\mathrm d}
\newcommand{\mi}{\mathrm i}
\newcommand{\vphi}{\varphi}
\newcommand{\uvc}[1]{\boldsymbol{\mathrm{\hat #1}}}

\begin{document}
	\title{Self-organized flows in phase-synchronizing active fluids}
	\author{Brato Chakrabarti}\affiliation{Center for Computational Biology, Flatiron Institute, New York, NY 10010, USA}
	\author{Michael J. Shelley}\affiliation{Center for Computational Biology, Flatiron Institute, New York, NY 10010, USA} 
	\affiliation{Courant Institute, New York University, New York, NY 10012, USA}
	\author{Sebastian Fürthauer}	\email[Email address: ]{fuerthauer@iap.tuwien.ac.at }\affiliation{Center for Computational Biology, Flatiron Institute, New York, NY 10010, USA}
	\affiliation{Institute for Applied Physics, TU Wien, A-1040 Wien, Austria}
	\date{\today}
	
	\begin{abstract}	
	Many active biological particles, such as swimming microorganisms or motor-proteins, do work on their environment by going though a periodic sequence of shapes. Interactions between particles can lead to the phase-synchronization of their duty cycles. Here we consider collective  dynamics in a suspension of such active particles coupled through hydrodynamics.
	We demonstrate that the emergent non-equilibrium states feature stationary patterned flows and robust unidirectional pumping states under confinement. Moreover the phase-synchronized state of the suspension exhibits spatially robust chimera patterns in which synchronized and phase-isotropic regions coexist within the same system. These findings demonstrate a new route to pattern formation and could guide the design of new active materials.
	\end{abstract}
	\pacs{...}
	\maketitle
	
Many microbes, motor-laden cytoskeletal assemblies, and many of their non-living analogs do work on their surroundings by undergoing a mechano-chemical duty cycle 
\cite{saintillan2013active,marchetti2013hydrodynamics}. On timescales longer than the duty cycle, this often results in net cycle-averaged dipolar stresses, which if strong enough, drive instabilities resulting in local ordering  and collective motion \cite{saintillan2008instabilities,baskaran2009statistical}. In this Letter, we study suspensions of immotile active particles whose cycle-averaged force dipole is zero. For these, the above-mentioned route towards self-organization is precluded \cite{brotto2015spontaneous}. We report an alternate route to collective motion: suspensions of active particles can spontaneously form collectively moving states by synchronizing their phases. In confinement, this mechanism can produce steady unidirectional flows, where active particles self-organize into an active pump or oscillate collectively. Our findings reveal the existence of a previously unknown mode of collective motion for microbial, cytoskeletal, and engineered immotile active particle suspensions. 

We quantify the progression along an internal duty cycle of an active particle by a phase $\vphi \in [0,2\pi)$ that modulates the active stress. Indeed, flow measurements of swimming algae show that the dipole strength associated with their  motion changes sign coherently over the flagellar beat cycle \cite{guasto2010oscillatory}.
While most active matter theories are formulated in terms of cycle averaged stresses \cite{saintillan2008instabilities}, phase-coherent versions also exist \cite{furthauer2013phase,leoni2014synchronization,banerjee2017active,o2017oscillators}. Intriguingly, linear stability analysis of these theories suggests that active particles can spontaneously synchronize their phases and form patterned states. By analogy to another much-studied example of phase patterning, the formation of metachronal waves in ciliary arrays \cite{uchida2010synchronization,brumley2015metachronal,meng2021conditions,chakrabarti2021multiscale}, where ciliary coordination results in large scale fluid flows it has been speculated \cite{furthauer2013phase,leoni2014synchronization,banerjee2017active} that the phase-patterned states emerging in synchronizing active particles could also drive fluid flows. However, the actual emergent properties of phase-synchronized active matter have remained unknown. Here, to resolve this we study the non-equilibrium dynamics of phase-coherent active fluids \cite{furthauer2013phase} numerically and analytically.
We find that flows and patterns in phase-synchronizing active materials emerge via a mechanism that is distinct from the classic instabilities known for other active matter systems \cite{simha2002hydrodynamic,saintillan2008instabilities,marchetti2013hydrodynamics}.  

\begin{figure*}
	\centering
	\includegraphics[width=1\linewidth]{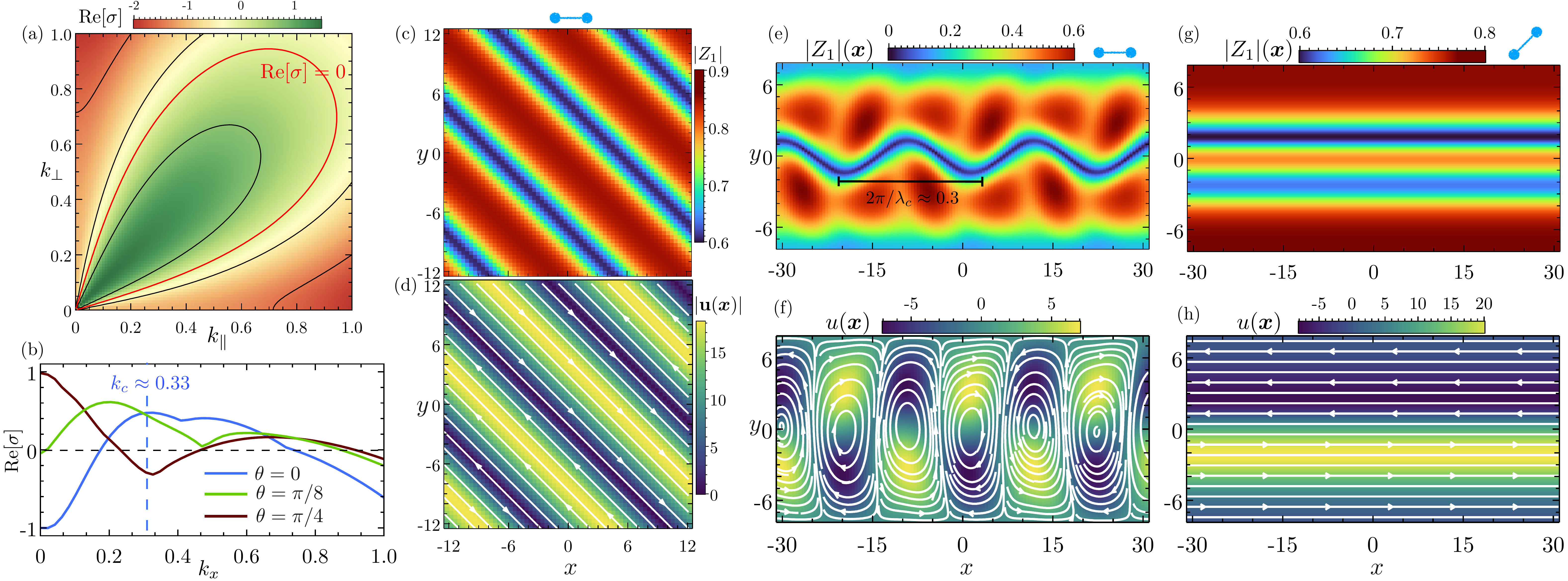}
	\caption{(a) The real part of the growth rate $\sigma$ as a function of wavenumbers $\{k_\pll,k_\perp\}$ in a periodic box, where $k_\pll$ and $k_\perp$ are wavevectors parallel and normal to the fixed particle orientation $\bn$, respectively. The unstable region is within the red curve.  (b) Growth rate computed from the linear stability as a function of the wavenumber $k_x$ for three different particle alignments.  The dominant wavenumber $k_c$ for $\theta = 0$ is indicated. A snapshot of the Kuramoto order parameter $|Z_1(\bx)|$ (top) and instantaneous streamlines (bottom) from the emergent traveling wave state is shown for a periodic box (c,d)  and in a channel for $\theta =0$ (e,f), and $\theta = \pi/4$ (g,h). Parameters: $\eta = 2, \gamma = 1, c_0 = \omega_0 = 8, X(\vphi) = \cos \vphi + 2 \sin \vphi$. The cartoon of the active particle indicates the fixed orientation of the particles. Channel height is chosen to be $H = 5\pi$.}
	\label{fig:fig1}
\end{figure*}

We start by developing a theory for active phase synchronizing suspensions, which extends \cite{furthauer2013phase} by including orientational dynamics.
Consider a 2D suspension of apolar active particles immersed in a Stokesian fluid. The probability density of  particles with phase $\vphi$ at position $\mathbf x$ is given by $\psi(\bx,\vphi,t)$ and obeys the  Smoluchowski equation
\begin{equation}\label{eq:smol}
	\partial_t \psi + v_\alpha \partial_\alpha \psi + \partial_\vphi (\dot{\vphi} \psi) - D \partial_\alpha^2 \psi = 0,
\end{equation}
where $D$ is the translational diffusion and $\bv(\bx) = (u,v)$ is the the fluid velocity. In principle each particle is characterized by an additional unit vector $\bn(\bx,t)$ that describes particle orientation. In this letter, we are only concerned with sharply aligned 2D nematic states  where the nematic tensor is given as $Q_{\alpha \beta} = n_\alpha n_\beta - \delta_{\alpha \beta}/2$. In this limit, particles at location $\bx$ simply point along the local director $\bn(\bx,t) = (\cos \theta(\bx,t),\sin \theta(\bx,t))$. 

Following the arguments in \cite{furthauer2013phase} we take the phase-velocity $\dot{\vphi}$ as
\begin{equation}\label{eq:omega}
	\dot{\vphi} = \Omega_0  + X(\vphi) Q_{\alpha \beta} u_{\alpha \beta} - d_\vphi \partial_\vphi \ln \psi,
\end{equation}
where $u_{\alpha \beta} = (\partial_\alpha v_\beta + \partial_\beta v_\alpha)/2$ is the rate-of-strain tensor and Einstein summation is implied. The phase velocity of isolated active particles in a quiescent fluid is $\Omega_0$. Fluid flows, which can be generated externally or by other particles, modify the phase dynamics. We make this contribution proportional to the energy dissipation $\mathcal{E} = Q_{\alpha \beta} u_{\alpha \beta}$ \cite{de2013non} by a single particle. The coupling function $X(\vphi)$ is phase dependent. This reflects that the same applied strain rate $u_{\alpha \beta}$ that helps a particle's extension in its extensile phase, will hinder its contraction during its contractile phase. The last term in Eq.~\eqref{eq:omega} accounts for phase-diffusion.

The driven incompressible Stokes equation describes the motion of the fluid in which particles are immersed
\begin{align}
	\eta \Delta \bv - \nabla q + \nabla \cdot \left(\siga + \sige\right) &= \mathbf{0}, \label{eq:stokes}
\end{align}
where  $\eta$ is the shear viscosity, and $q$ is the pressure that is determined by the incompressibility condition $\nabla \cdot \bv = 0$. Further,
$\siga$ is the active stress that the particles exert on the fluid. The elastic stress $\sige$ is generated by alignment interactions between particles. The mechano-chemical duty-cycle of the particles gives rise to phase-dependent force dipoles. The active stress in the suspension  is given by
\begin{equation}\label{eq:stress}
	\sigma^{\text{act}}_{\alpha \beta}(\bx,t) = Q_{\alpha \beta}  \int_0^{2 \pi} \psi(\bx,\vphi,t) s(\vphi) \dot{\vphi}\ \md \vphi,
\end{equation}
where $s(\vphi)$ is a $2 \pi$-periodic function that describes the details of the force production through the duty cycle. In this letter, we assume that the active stresses are generated through cyclic surface deformation of the particles and thus proportional to the phase-speed $\dot{\vphi}$. We further choose  $s(\vphi) = S\cos \vphi$ and thus particles with a constant phase-velocity generate no cycle-averaged active stress. 

Particle-particle alignment generally produces an elastic stress $\sigma_{\alpha \beta}^{\text{\text{el}}}(\bx,t) = n_\alpha h_\beta$ on the fluid, where the molecular alignment field $\mathbf h=-\delta f^{\text{el}}/{\delta \mathbf n}$, and $f^{\text{el}}$ is the free energy density associated with deformations of the nematic alignment field \cite{de1993physics}. With this, the evolution of the director field for rod-like bodies obeys
\begin{equation}\label{eq:polar}
	\frac{\mathrm{D} n_\alpha}{\mathrm{D}t} = \frac{1}{\gamma}h_\alpha - (\partial_\alpha v_\beta) n_\beta,
\end{equation}
where $\mathrm{D}/\mathrm{D}t = \partial_t + v_\alpha \partial_\alpha$ is the material derivative, $\gamma$ is the rotational viscosity, and $\partial_\alpha v_\beta$ is the  velocity gradient tensor. Here we use single constant Frank elasticity approximation to write $f^{\text{el}}=K (\nabla \bn) : (\nabla \bn)/2 + \xi (\bn \cdot \bn -1)/2$, where $\xi$ is the Lagrange multiplier that enforces the unit length of the director \cite{de1993physics}. The interaction of elasticity with activity introduces the Freedericksz length  $\ell_\text{\text{f}} = L \sqrt{K/S\Omega_0}$, where $L$ is the system size \cite{alert2020universal}. In the limit, $\ell_f \gg L$, the director field does not deform, and the molecular alignment field, $h_\alpha = \gamma  n_\beta \partial_\alpha v_\beta$, acts as the Lagrange multiplier that enforces $\mathrm{D} \bn/\mathrm{D}t = 0$.



To describe synchronization in a continuous system, we introduce the order-parameter fields $Z_n(\bx,t) = \int_0^{2 \pi} e^{\mi n \vphi }\psi(\bx,\vphi,t) \ \md \vphi$
where $n \in \mathbb{Z}$ i.e. the Fourier coefficients of the distribution function in $\vphi$ \cite{furthauer2013phase}. By construction $Z_0 \equiv c(\bx,t)$ is the concentration field and $Z_1(\bx,t)$ is the complex Kuramoto order field. Synchronized states in the suspension corresponds to $|Z_1| \approx 1$, while $|Z_1| \approx 0$ is associated with phase-disorder. The   active stress in Eq.~\eqref{eq:stress} can now be written as $\sigma_{\alpha \beta}^{\text{act}}  = \mathcal{M}(\bx) Q_{\alpha \beta}$, where
\begin{equation}\label{eq:strmom}
	\mathcal{M}(\bx) = \Omega_0 s_{m} Z_{m}+  u_{\mu \nu} Q_{\mu \nu} s_{m} X_{n} Z_{n+m}  -\mathrm{i} m d_\vphi s_{m} Z_{m}.
\end{equation}
Here $s_m,X_m$ are the Fourier coefficients of $s(\vphi)$ and $X(\vphi)$, respectively. 
Using Eq.~\eqref{eq:smol} we obtain
\begin{equation}\label{eq:Zeq}
	\frac{D Z_n}{D t} - D \Delta Z_n + n^2 d_\vphi Z_n = \mathrm{i}n \left(\Omega_0 Z_n +u_{\alpha \beta} Q_{\alpha \beta}X_{m}  Z_{n+m} \right) 
\end{equation}
The above evolution for $Z_n$ yields a hierarchy of coupled equations involving higher-order moments. If $d_\vphi > 0$ the higher-order moments decay rapidly as seen from the left hand side of Eq.~\eqref{eq:Zeq}. Here we evolve $Z_{1,2,3}$ and approximate $Z_4 \approx Z_2^2/c$.  Retaining higher order moments did not alter the results obtained through this closure. Limited simulations of the full kinetic model also provided similar results. Hence, here we will  explore the model by evolving Eqs.~\eqref{eq:stokes},\eqref{eq:polar},\eqref{eq:strmom}, and \eqref{eq:Zeq}. As outlined in \cite{furthauer2013phase}, the model exhibits hydrodynamic instabilities when the coupling function $X(\vphi)$ is phase-shifted from $s(\vphi)$. So we use $X(\vphi) = A \cos(\phi - \bar{\phi})$, where $\bar{\phi}$ is the phase-shift. Including higher harmonics in $X(\vphi)$ did not alter the qualitative behavior of the system.

\begin{figure*}
	\centering
	\includegraphics[width=1\linewidth]{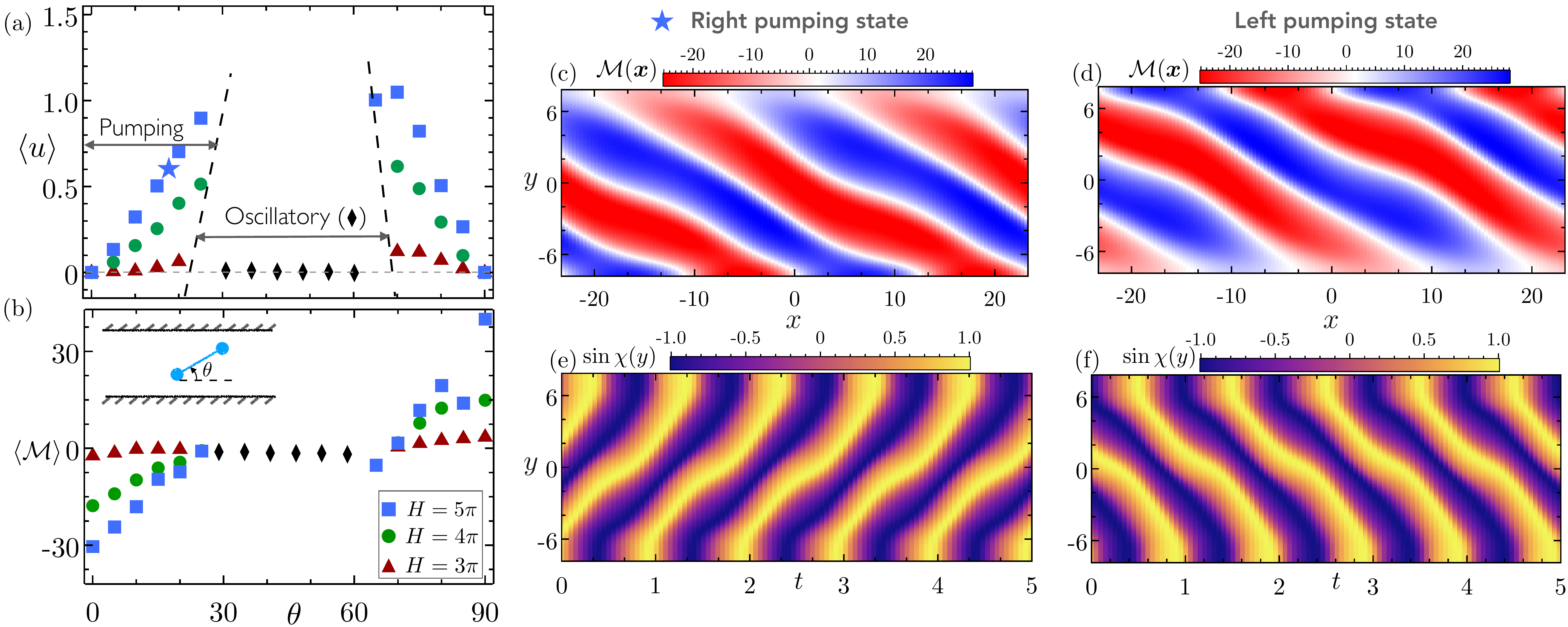}
	\caption{(a) Phase diagram of the two emergent states: spontaneous pumping and oscillations and (b) the mean amplitude of the active stress $\langle \mathcal{M}\rangle$ as a function of $\theta$ for varying channel height $H$. (c,d) Variation of $\mathcal{M}(\bx)$ in the channel demonstrate that through synchronization the suspension behaves as an extensile (or contractile) fluid. The direction of pumping results from the spatial distribution of the stress. (e,f). Variation of the phase along the channel width as a function of time illustrates the existence of metachronal phase waves. Parameters: $H = 5 \pi$, $L = 3 H$, $\theta = \pi/8$, $\gamma = 1$, $\eta = 2$, $\omega_0 = 8$, $d_\vphi = 1$, $D = 1, X(\vphi) = \cos \vphi + 2 \sin \vphi$.}
	\label{fig:fig2}
\end{figure*}

\begin{figure*}
	\centering
	\includegraphics[width=1\linewidth]{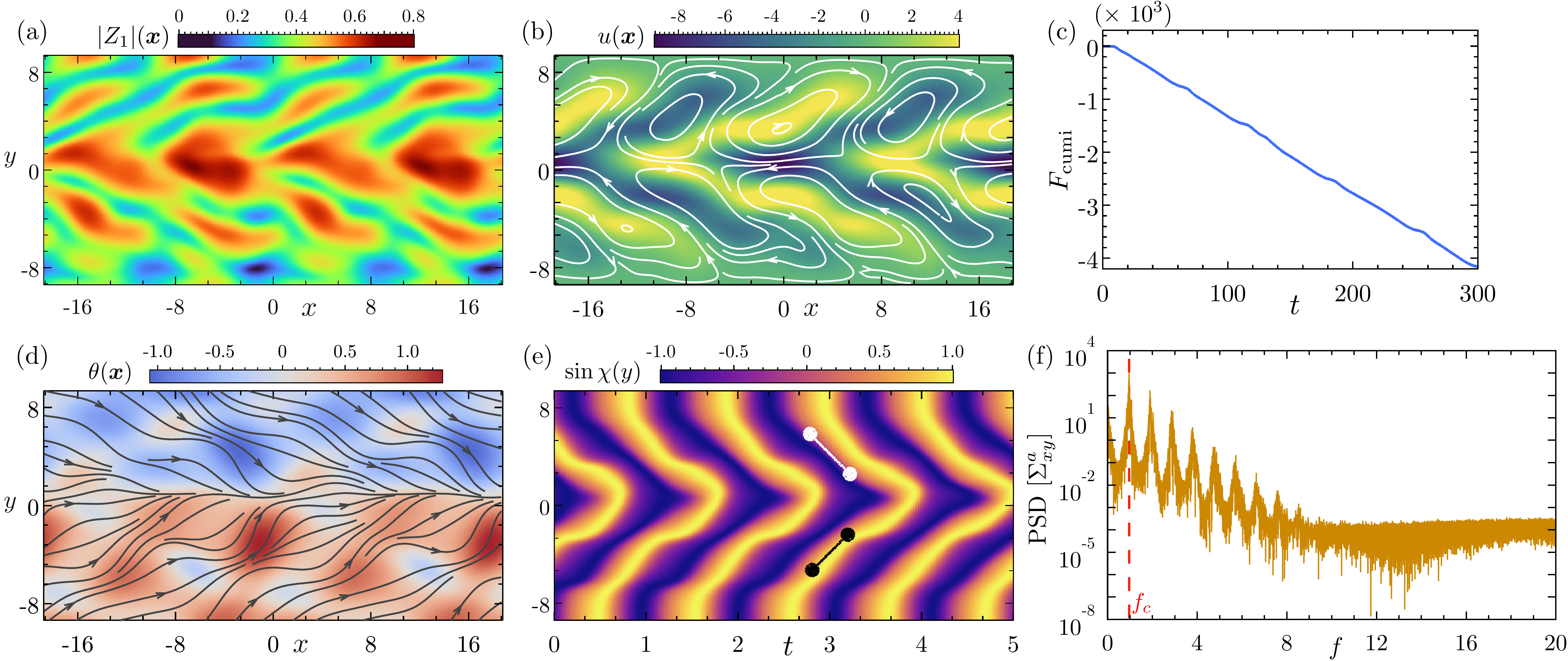}
	\caption{(a) The Kuramoto order parameter from the final state of traveling waves illustrate the emergence of Chimera states with regions of asynchrony $|Z_1| \approx 0$ and synchrony $|Z_1| \approx 1$. (b) Associated snapshot of the velocity field and the streamlines highlight the formation of vortical structures and fluid jets.  (c) The cumulative flux $F_{\text{cumi}}(t) = \int_0^t \langle u \rangle (t') \md t'$ demonstrates that the coupled phase-alignment dynamics result  in steady unidirectional pumping. (d) $\theta(\bx)$ showcases the variation in the orientation and the associated director field. (e) Phase kymograph from the state of pumping showcases the existence of metachronal waves that is behind the spontaneous pumping. (f) Power spectral density of the steady-state active shear stress. We see peaks at different harmonics, with the strongest one highlighted in red. Prameters: $H = 6 \pi, L = 12 \pi,\eta = 2, \gamma = K = 0.15,  c_0 = \omega_0 = 8$.}
	\label{fig:fig3}
\end{figure*}


\emph{Linear stability:} This is a complicated dynamical system. To delineate the role of synchronization phenomena, we first seek understanding of the relatively simpler case, where the particles have fixed orientation  ($\dot{n} = 0$), relevant in the limit of $\ell_\text{f} \gg L$.
For $d_\vphi > 0$ the phase-disordered state, with $Z_0 = c_0$, $\bv = \mathbf{0}$, and $Z_n = 0$, is the sole fixed point of this problem. For periodic boundary conditions,
consistent with \cite{furthauer2013phase}, the phase-disordered state has a long wave-length Hopf bifurcation. We find that the most unstable wavevector in this problem is set by the direction of largest shear. Due to incompressibility this is at an angle of $45^\circ$ above or below the vector $\mathbf{n}$; see  Fig.~\ref{fig:fig1}(a). We next analyze the stability inside a channel of height $H$ with no-slip applied at $y = \pm H/2$. We set $d_\vphi \partial_y Z_n|_{y=\pm H/2} =0$ that enforces impermeable boundaries. With the Ansatz $\{\bv,Z_1\} \rightarrow \varepsilon \{\bv(y),Z_1(y)\} \exp(\mi k_x + \sigma t)$, where $\varepsilon \ll 1$, the growth rate $\sigma$ can be computed numerically; see Fig.~\ref{fig:fig1}(d). Like in the periodic domain, the phase-disordered state is linearly unstable. Here, the wavenumber of the fastest growth $k_c$ is set by the channel height $H$ and the particles' orientation. We next study the dynamics of the system beyond the linear regime using spectral simulations \cite{burns2020dedalus}.

\emph{Nonlinear dynamics:} Figures~\ref{fig:fig1}(c,d) and (e,f) show the spatial structure of the phase-order parameter and the associated fluid flows for fixed $\bn = \uvc{e}_x$  inside a periodic box and a channel, respectively. In periodic domains, the Kuramoto order parameter $|Z_1|(\bx)$ and consequently the velocity $\bv(\bx)$ settle into a striped traveling wave (see movie S1). The patterns are at the scale of the computational box, consistent with having emerged from a long-wavelength instability (Fig.~\ref{fig:fig1}(c,d)). Indeed we chose the parameters which made the fundamental scale of the box the fastest growing wavelength. Further, again consistent with the stability analysis, the modulated stripes are orthogonal to the wave-vector of maximum growth ($45^\circ$) and the wave-speed is parallel to it. This means that the emergent state is an example of an orthoplectic wave \cite{han2018spontaneous}.

In channels, with $\bn = \uvc{e}_x$, Eq.~\eqref{eq:omega} at the boundaries reduces to $\dot{\vphi} = \Omega_0 - d_\vphi \partial_\vphi \ln \psi$. This  predicts that $|Z_1| \approx 0$ near the wall, which is indeed seen from the simulations on Fig.~\ref{fig:fig1}(e).  Consequently, the instability produces a traveling wave along $\uvc{e}_x$ with a modulated  pattern in $|Z_1|$, which again has a width of  $\lambda_c \approx 2 \pi/k_c$ consistent with linear stability. Figure.~\ref{fig:fig1}(e) further shows that the emergent state is a Chimera state \cite{abrams2004chimera} with the coexistence of the phase-disordered ($|Z_1| \approx 0$) and phase-synchronized ($|Z_1| \simeq 1$) regions. The associated fluid flows  show the formation of vortical structures between them (see Fig.~\ref{fig:fig1}(f) and movie S2). Our results demonstrate that even when the system size is below the Freedericksz length, phase-synchronization can provide a novel route towards instabilities. The associated self-organized flows seen in Fig.~\ref{fig:fig1} are distinct from those seen in classical orientation instabilities \cite{simha2002hydrodynamic,saintillan2008instabilities,marchetti2013hydrodynamics,alert2020universal,alert2021active}, which all depend on non-uniform alignment throughout the domain.

We next \textcolor{black}{characterize the} emergent states in the channel. For this, we first probe the zero wavenumber limit associated with transport. We find from Eq.~\eqref{eq:stokes} that $\langle u \rangle = 0$ when $\mathbf n = \hat{\mathbf{e}}_x$ or $\uvc{e}_z$; where $\langle \cdot \rangle$ denotes the spatial average of any field variable. \textcolor{black}{As we increase $\theta$ starting from $\theta=0$, the system instead settles into a pumping state. For these, $|Z_1|$ evolves into a traveling wave along the channel length accompanied by an emergence of a steady flow along $\uvc{e}_x$. The associated fluid flow is characterized by tilted vortices and fluid jets (see movie S3). As can be seen from Fig.~\ref{fig:fig2}(a), the steady flux $\langle u \rangle$ from the final state increases with the channel height $H$ and the particle orientation $\theta$. The symmetry of the problem means that pumping fluid in either direction is equally likely, and initial conditions choose the final direction.}

\textcolor{black}{To better understand the underlying mechanism of the pumping, we next study $ \mathcal{M}(\bx)$ from the final state, which is the amplitude of the active stress $\sigma_{\alpha \beta}^{\text{act}} = \mathcal{M} Q_{\alpha \beta}$. As shown  in Fig.~\ref{fig:fig2}(c,d), $\langle \mathcal{M}(\bx) \rangle$ is identical for both left and right pumping  states with the same $\theta$ and has a non-zero mean throughout the pumping regime; see Fig.~\ref{fig:fig2}(b). This means, that the system has settled into a state of mean extensile (or contractile) active stress (see Fig.~\ref{fig:fig2}(b)) through phase-synchronization even though the constituent active particles do not have any mean dipole. The direction of pumping is dictated by the difference of the values of $\mathcal{M}(\bx)$ at the boundaries.  Simply put, the axial force per unit depth exerted by the active stresses on the walls is $\int \md x  [(\mathcal M \mathcal Q_{xy})|_{y=H/2}-(\mathcal M \mathcal Q_{xy})|_{y=-H/2}]$. This is nonzero if the phase ordering is different for the particles on opposite walls and $\theta \neq 0,\pi/2$. In this case, the particles on average act more pusher like in one wall and puller like near the other. This mechanism is further illustrated by the  phase kymographs shown on Fig.~\ref{fig:fig2}(e,f), where we plot the phase $\chi(\bx,t) = \mathrm{arg}[Z_1 (\bx,t)]$ at a given $x$ as a function of time. Figure~\ref{fig:fig2}(e,f) indicates  the existence of left and right traveling metachronal phase waves associated with the two different directions of pumping. Thus, our simulations demonstrate  (i) the spontaneous emergence of coherent fluid transport by phase synchronization, and (ii) the spontaneous emergence of polar pumping states from a globally aligned suspension.}

\textcolor{black}{We note that this pumping state loses stability as $\theta$ approaches $\pi/4$, giving rise to oscillatory states instead. Numerical evidence on Fig.~\ref{fig:fig2}(b) suggests that the loss of stability of the pumping states is associated with the suspension transitioning from being extensile to contractile with $\langle \mathcal{M} \rangle \approx 0$. In the SI we discuss more on the structure of the bifurcation branches. Figure~\ref{fig:fig1}(g,h) shows an oscillatory state for $\theta =\pi/4$ (see movie S3). The snapshots highlight that the phase modulations span the entire channel width, consistent with the linear stability; see Fig.~\ref{fig:fig1}(b). The Kuramoto order parameter and the associated velocity field organize into traveling waves moving from wall to wall and resulting in the formation of fluid jets along the channel length whose direction oscillate periodically over time. Importantly, since these waves do not break translational symmetry in the $\uvc{e}_x$ direction when averaged over a period, they generate no mean flow. As can be seen from Fig.~\ref{fig:fig2}(a), such system wide oscillatory states exist around a range of orientation around $\theta=\pi/4$. At high enough $\theta$, the pumping state re-emerges.}



We next asked how the system self-organizes if the director $\mathbf n$ evolves according to Eq.~\eqref{eq:polar} ($\dot{n} \neq 0$). In Fig.~\ref{fig:fig3} we show simulation results in a channel where we  set $\partial_y \theta|_{y = \pm H/2} = 0$. \textcolor{black}{We observe that for all the explored channel heights, the active fluid always settles into a traveling wave state or a periodic orbit (see movie S5), as evident from the multiple harmonics in the power spectral density of the active shear stress; see Fig.~\ref{fig:fig3}(f). This emergent state is characterized by coherent fluid pumping and propagating metachronal phase waves as seen from the phase-kymograph at $x=0$; see Fig.~\ref{fig:fig3}(c,e). Figure~\ref{fig:fig3}(d) shows the director field that develops in the channel. Both the orientations and the phases of the particles on average, remain mirror-symmetric around $y=0$. The average particle orientations on either side of the channel centerline are shown schematically on Fig.~\ref{fig:fig3}(e). The left and right-going metachronal phase waves (on top and bottom half of the channel) and the associated average alignment of the particles aid fluid pumping resulting in coherent transport. } 


We finally probe whether the emergent fluid flows in the $\dot{n} \neq 0$ system result from the dynamics of particle orientation or synchronization. For this, we study the dynamics of a $\dot{n} \neq 0$ nematic fluid driven by an active stress $\sigma_{\alpha \beta}^{\text{act}} = \zeta_0 \sin(2 \pi f_c t) n_\alpha n_\beta$. Here, $f_c$ is the frequency of the forcing and is obtained from the power spectral density of the steady-state active shear stress $\sigma_{xy}^{\text{act}}$ of the active fluid; see Fig.~\ref{fig:fig3}(f). The amplitude $\zeta_0$ is chosen to be the maximum of the measured $\sigma_{xy}^{\text{act}}$. This active forcing mimics the stresses generated by oscillating dipoles that are free to align but unable to synchronize. We find, that this stress does not drive any alignment instability, and the fluid velocity decays after a short transient. At the heart of this is a separation of time scales associated with the relaxation time of the director field $\gamma L^2/K$ and the time scale of the forcing $\Omega_c \equiv 2 \pi f_c$. For the present example, $\Omega_c \gg \gamma L^2/K$; as a result, on average, the fluid does not experience any induced activity and relaxes to a homogeneous state with no flow. This shows that synchronization drives the alignment instabilities and the flows shown in Fig.~\ref{fig:fig3}. 

To conclude, we have uncovered a new route towards self-organized flows in active suspensions. Rather than stemming from the cycle averaged stress dipoles, which active particles exert on a time scale much larger than \textcolor{black}{their duty cycles, this second route stems from synchronization between particles. This synchronization phenomena can give rise to coherent transport under confinement even when the cycle averaged stress is zero for the constituent particles.
The emergent traveling waves are Chimeras \cite{martens2013chimera} with regions of high and low synchrony. }

\textcolor{black}{It is known that coupled mechano-chemical oscillators are involved in important processes like somitogenesis \cite{uriu2017framework} and have also been invoked to explain bacterial pattern formation \cite{igoshin2001pattern}. However, in these examples, the coupling between oscillators happen through chemical degrees of freedom \cite{thutupalli2011swarming}. The mechanism we uncovered here is based on hydrodynamic interactions and provides a novel route to phase patterning in mechanochemical oscillators that is 
clearly distinct from \cite{uriu2017framework,igoshin2001pattern,o2017oscillators}. It is also distinct from other mechanisms that have been invoked to explain biological patterns such as the classical Turing instability \cite{turing1990chemical} and its generalizations to phase-insensitive active matter \cite{howard2011turing,bois2011pattern}. In \textit{toto}, the synchronization provides a new route towards emergent dynamics in active materials. This finding will inform our understanding of collectively moving microbes, motorized cytoskeletal structures, and potentially facilitate the engineering of active materials \cite{thutupalli2011swarming} that can act as micropumps.}




{\bf Acknowledgments:}
S.F. is supported by the Vienna Science and Technology Fund (WWTF) and the City of Vienna through project VRG20-002. MJS acknowledges support by the National Science Foundation under awards DMR- 1420073 (NYU MRSEC) and DMR-2004469.

\bibliography{bibfile}

\end{document}